\documentclass{article}	
\usepackage{multicol}
\usepackage{graphicx}
\usepackage{color}
\usepackage{amsmath}
\usepackage[margin=0.5in]{geometry}
\usepackage{float}
\usepackage{cite}
\usepackage{MnSymbol}
\usepackage{authblk}
\definecolor{mycolor1}{rgb}{0.3, 0.9, 0.9}
\definecolor{mycolor2}{rgb}{0.1, 0.7, 0.3}
\newcommand{\subfigimg}[3][,]{%
  \setbox1=\hbox{\includegraphics[#1]{#3}}% Store image in box
  \leavevmode\rlap{\usebox1}% Print image
  \rlap{\hspace*{10pt}\raisebox{\dimexpr\ht1-1\baselineskip}{#2}}% Print label
  \phantom{\usebox1}% Insert appropriate spcing
}
\begin{document}
\title{Topographical coloured plasmonic coins}

%% Notice placement of commas and superscripts and use of &
%% in the author list
\author[1,2,4,*]{J-M Guay}
\author[1,2]{A. Cal\`a Lesina}
\author[1,2,4]{G. C$\hat{o}$t$\acute{e}$}
\author[1]{M. Charron}
\author[1,2]{L. Ramunno}
\author[1,2,3]{P. Berini}
\author[1,2,4]{A. Weck}
%\author{J-M Guay$^{1,2,4}$, A. Cal\`a Lesina$^{1,2}$, G. C$\hat{o}$t$\acute{e}^{1,2,4}$, M. Charron$^{1}$, L. Ramunno$^{1,2}$, P. Berini$^{1,2,3}$ and A. Weck$^{1,2,4}$}
\affil[1]{Department of Physics, University of Ottawa, ON, K1N 6N5, Canada}
\affil[2]{Centre for Research in Photonics, University of Ottawa, ON, K1N 6N5, Canada}
\affil[3]{School of Electrical Engineering and Computer Science, University of Ottawa, ON, K1N 6N5, Canada}
\affil[4]{Department of Mechanical Engineering, University of Ottawa, ON, K1N 6N5, Canada}
\affil[*]{email: jguay036@uottawa.ca}
\maketitle

\begin{abstract}
The use of metal nanostructures for colourization has attracted a great deal of interest with the recent developments in plasmonics. However, the current top-down colourization methods based on plasmonic concepts are tedious and time consuming, and thus unviable for large-scale industrial applications. Here we show a bottom-up approach where, upon picosecond laser exposure, a full colour palette independent of viewing angle can be created on noble metals. We show that colours are related to a single laser processing parameter, the total accumulated fluence, which makes this process suitable for high throughput industrial applications. Statistical image analyses of the laser irradiated surfaces reveal various distributions of nanoparticle sizes which control colour. Quantitative comparisons between experiments and large-scale finite-difference time-domain computations, demonstrate that colours are produced by selective absorption phenomena in heterogeneous nanoclusters. Plasmonic cluster resonances are thus found to play the key role in colour formation.
\end{abstract}

\begin{multicols}{2}
\section{Introduction}
Metal nanoparticles (NPs) are used in a multitude of applications, but perhaps the oldest is as a colourizing agent when dispersed in a host dielectric, as in the (dichroic) Lycurgus cup\cite{Freestone2007,Barch2015}. Exposed to optical radiation, metal NPs exhibit scattering properties due to excited plasmons, which depend on their shape, size, composition and the host medium\cite{Mie,Doyle1989,Murray2007,Liz2006}.\\
Producing colours by exploiting plasmonic effects on metal nanostructures is of interest because the colours can last a long time (e.g., the Lycurgus cup), can be rendered down to the diffraction limit\cite{Kumar2012}, and can be used in any metal colouring or marking application where inks, paints or pigments should be avoided for environmental, health, cost or other reasons.
\\ 
Fabrication techniques for plasmonic colouring of metal surfaces include laser interference lithography (LIL)\cite{Gallinet2015}, electron beam lithography (EBL)\cite{Roberts2014,Tan2014}, ion beam lithography (IBL) or milling (IBM)\cite{Cheng2015}, and hot embossing or nanoimprint lithography (NIL) \cite{Gallinet2015,Clausen2014}. Kumar et al.\cite{Kumar2012} fabricated coloured images via EBL having a resolution as high as ≈ 100 000 pixels/inch$^{2}$. However, the production of large coloured surfaces is incompatible with the demands of low-cost mass-manufacturing. Furthermore, such processes generally require flat surfaces, and with the exception of periodic structures below the diffraction limit of visible light \cite{Clausen2014,Wu2013}, colours produced are generally angle-dependent.
Femtosecond (fs) lasers have been considered ideal for metal colourization due to their ablation characteristics, e.g., the tendency to preferentially release NPs, compared to the large clusters and chunks produced by nanosecond pulses \cite{Tillack2004,Perriere2007,Balling2013}. Guo et al. \cite{Vorobyev2013a} showed that exposing metals to fs pulses could yield highly absorptive surfaces (so called ‘black metals’), or surfaces producing a specific colour. However, only one set of laser parameters was reported for each colour \cite{Fan2014,Vorobyev2008,Fan2013b}, the colour palette was limited and, due to the underlying regular structure, in the case of femtosecond lasers, were angle-dependent \cite{Vorobyev2008c}. In addition, the low pulse energy of fs lasers restricts their use to low repetition rates, making the colouring process very time-consuming.
 
Picosecond (ps) lasers have lower costs and higher pulse energies. Fan et al. showed a limited colour palette on copper and each colour was associated to one set of laser parameters \cite{Fan2014}. Thermal effects for ps pulses are likely to play a role in the creation of colours, unlike femtosecond lasers, whose pulses are shorter than the thermal expansion time \cite{Nolte1997}.
Here, we report on a universal, high-throughput and deterministic process for producing a complete angle-independent colour palette composed of thousands of colours using a picosecond laser. This is achieved on unpolished noble metal surfaces, including surfaces with frosting and millimetre-scale topographical features. We demonstrate the process by colouring silver coins produced at the Royal Canadian Mint, set to be released to the general public at the end of 2016. Scanning electron microscopy (SEM) reveals that a full colour palette can be obtained by controlling the particle density on the surface.
Our experiments demonstrate that a very large set of laser parameter combinations can produce a given colour, as long as the total accumulated fluence remains the same, making the process scalable and time-efficient. 
\begin{figure}[H]
\centering
\begin{tabular}{@{}p{1\linewidth}@{}}
\subfigimg[width=\linewidth]{\textbf{\textcolor{white}{}}}{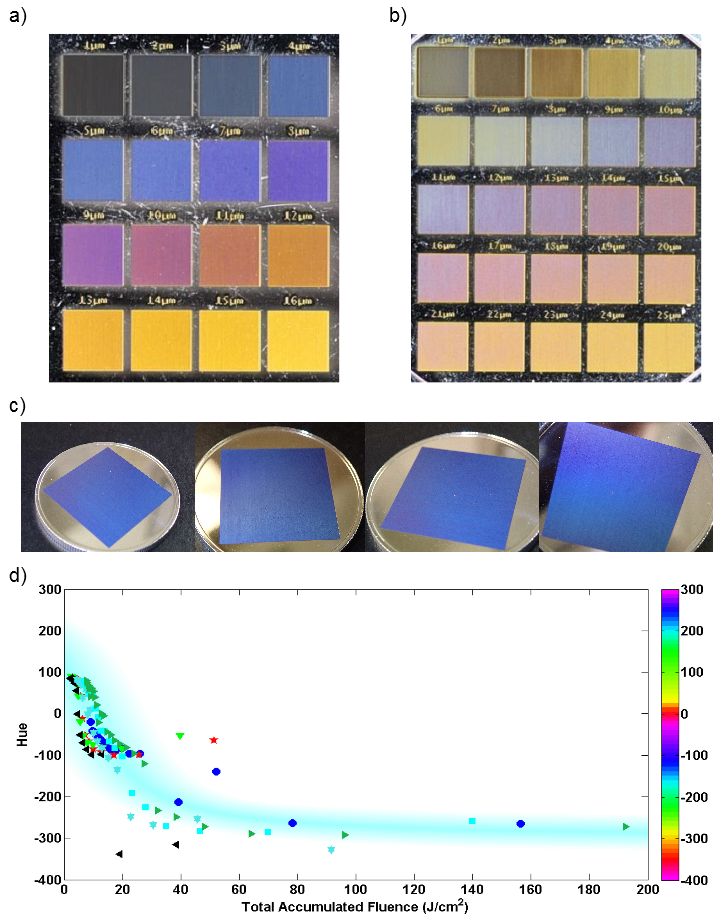}\\
\subfigimg[width=\linewidth]{\textbf{\textcolor{white}{}}}{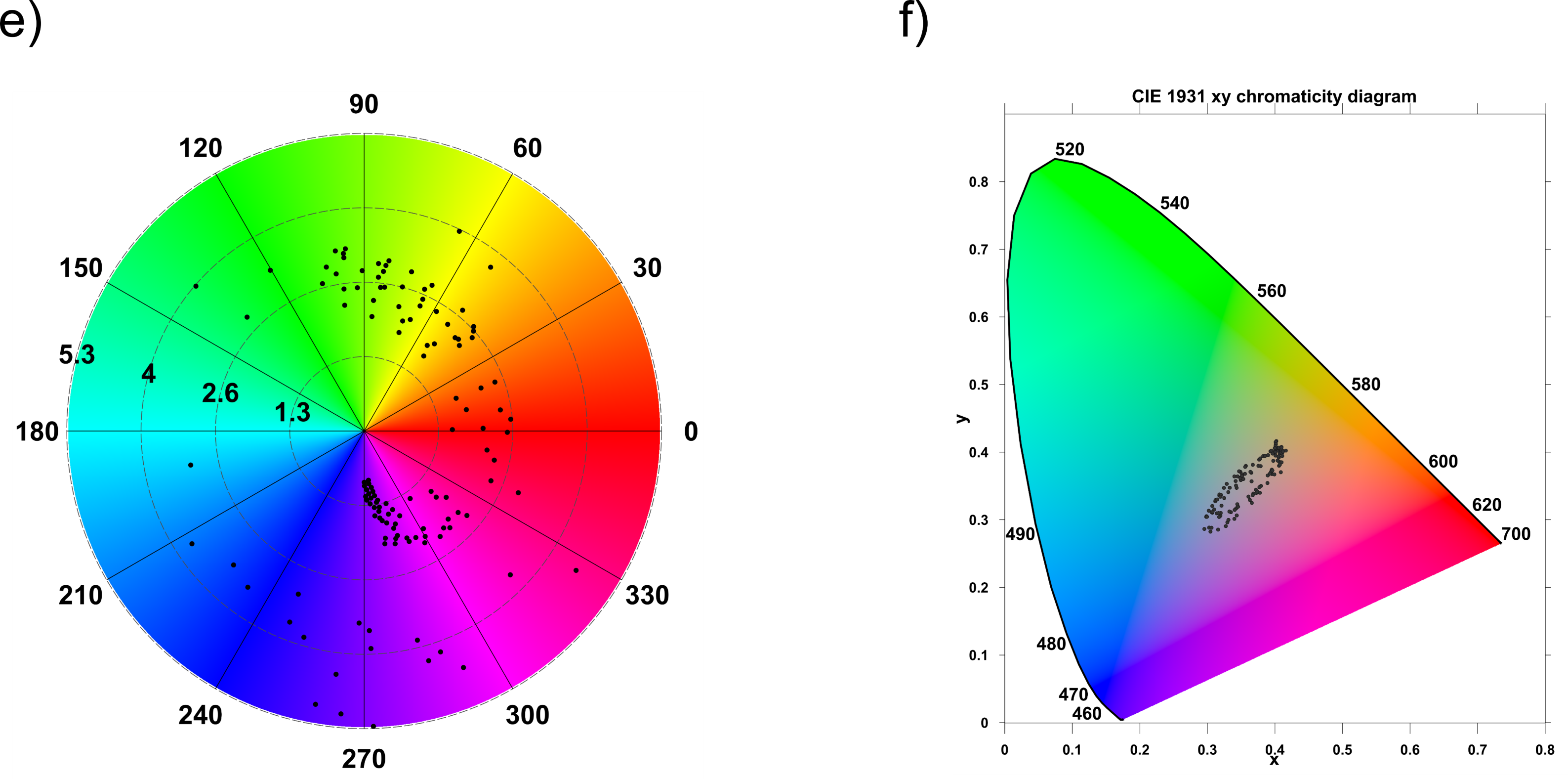}
\end{tabular}
\caption{25 mm$^2$ coloured squares obtained by 1 $\mu$m changing $L_s$  (marked above) for (a) a laser fluence of $\phi=1.67$ J/cm$^{2}$ with a laser marking speed of $v=50$ mm/s, and (b) $\phi=29.75$ J/cm$^{2}$ at $v=1000$ mm/s. (c) Photograph of blue angle-independent colouring of a flat silver coin observed at various angles. (d) Hue versus total accumulated fluence for ($\begingroup\color{blue}\bullet\endgroup$) $\phi=1.12$ J/cm$^{2}$ at $v=11$ mm/s, ($\begingroup\color{red}\star\endgroup$) $\phi=1.67$ J/cm$^{2}$ at $v=50$ mm/s, ($\begingroup\color{green}\filledtriangledown\endgroup$) $\phi=2.59$ J/cm$^{2}$ at $v=100$ mm/s, ($\begingroup\color{black}\filledtriangleleft\endgroup$) $\phi=6.26$ J/cm$^{2}$ at $v=250$ mm/s, ($\begingroup\color{mycolor1}\pentagram\endgroup$) $\phi=13.23$ J/cm$^{2}$ at $v=500$ mm/s, ($\begingroup\color{cyan}\filledsquare\endgroup$) $\phi=20.58$ J/cm$^2$ at $v=750$ mm/s, and ($\begingroup\color{mycolor2}\filledtriangleright\endgroup$) $\phi=29.75$ J/cm$^2$ at $v=1000$ mm/s; Hue can be seen to make a full 360$^{\circ}$ rotation (dashed line). (e) Polar plot representation of (d) with Hue plotted azimuthally and the logarithm of the total accumulated fluence plotted radially; a full rotation in Hue can be observed with increasing total accumulated fluence.  (f) CIE diagram of all the colors obtained using the different laser parameters from (d). }
\label{fig1}
\end{figure}
To theoretically understand the colour formation, we used large-scale computational electrodynamics to simulate the scattering from periodic distributions of different sized nanoparticles, with geometrical parameters based on statistical analysis of SEM images. Simulations show that several NP arrangements can produce the same colour, similar to what is observed experimentally, and that narrow band absorption due to plasmonic resonances in heterogeneous nanoclusters are critical for colour production.

\section{Results and Discussion}
\subsection{Angle-independent laser colouring of silver}
The exposure of pure silver to different laser parameters is observed to produce vivid angle-independent colours, Fig. 1 (a,b,c). The same process was used to produce similar colour palettes on copper and gold. These colours are highly reproducible and found to depend on the density of NPs covering the metal surface. 
\\
During laser ablation by raster scanning the sample surface, the number of particles re-deposited and accumulated from one line to the next is dictated by the inter-line spacing $L_s$, i.e., the spacing between two consecutive laser lines. The change in particle density with increasing distance from the laser ablated line can be seen in the supplementary material, Fig. S. 1.\\
Figs. 1 (a,b) show the colours produced on silver by raster scanning the surface, where different colours are produced by simply increasing Ls. In particular, increasing Ls from 1 to 12 $\mu$m with a laser marking speed of v = 50 mm/s results in a production rate of $\eta$ =0.05 to 0.8 mm$^2$/s (Fig. 1 (a)), while augmenting the laser marking speed to v = 1000 mm/s with Ls ranging from 1 to 25 $\mu$m (Fig. 1 (b)) results in production rates of $\eta$ =1 to 25 mm$^2$/s. The production rate is defined as $\eta$ = L$_s$v. The colour palettes in Figs. 1 (a,b) can be extended by simply reducing the line spacing between each successive line (i.e., increasing total accumulated fluence). The angle-independent colours produced on silver, Fig. 1 (c), cover the spectral and the nonspectral regions (e.g., Magenta) through a 360$^{\circ}$ rotation in Hue, Fig. 1 (d-f).
\\
The total accumulated fluence is defined as 
\begin{equation}
\Phi = \phi N_{eff} =\frac{a^{2}Ef}{vL_s},
\end{equation}
where the laser fluence is given by
\begin{equation}
\phi = \frac{2E}{\pi w_o^2},
\end{equation}
the number of effective laser shots is
\begin{equation}
N_{eff}= \underbrace{\sqrt{\frac{\pi}{2}}\frac{aw_of}{v}}_\text{intra-line}\underbrace{\sqrt{\frac{\pi}{2}}\frac{aw_o}{L_s}}_\text{inter-line},
\end{equation}
%corresponding to a production rate of
%\begin{equation}
%\eta = L_{s}v,
%\end{equation}
and $E$ is the laser pulse energy, ωo the beam waist radius, $f$ the laser repetition rate, and $a$ a correction factor due to the larger modified region when using a higher pulse energy. This correction factor was determined from semi-logarithmic plots used to determine the laser spot size, following the procedure detailed in \cite{Jandeleit1996}. The intra-line component of $N_{eff}$ encompasses the shot overlap within a local region, where $f/v$ is the distance traveled between successive laser pulses in a single laser line. The inter-line component, in comparison, considers the geometrical overlap, $a\omega_o/L_s$, between successive laser lines. At large spacings (\textit{i.e.}, lower total accumulated fluence and $a\omega_o/L_s$ $\rightarrow$ 1) the colours are observed to converge to yellow due to the absence of overlap between consecutive laser lines. Further evidence of this accumulation process can be observed from the absence of colours, other than yellow, in the last line of each of the coloured squares. The last line does not undergo the process of particle accumulation as there are no subsequent lasered lines. In addition, in the absence of overlap and for large distances between successive lines, LIPSS structures can be observed to form next to the laser ablated lines causing the yellow colors to be angle-dependent. In the overlap case no LIPSS structures are created or they are destroyed and the surface remains relatively smooth, Fig. S. 2 in supplementary materials. 
\begin{figure}[H]
\centering
\begin{tabular}{@{}p{1\linewidth}@{}}
\subfigimg[width=\linewidth]{\textbf{\textcolor{white}{}}}{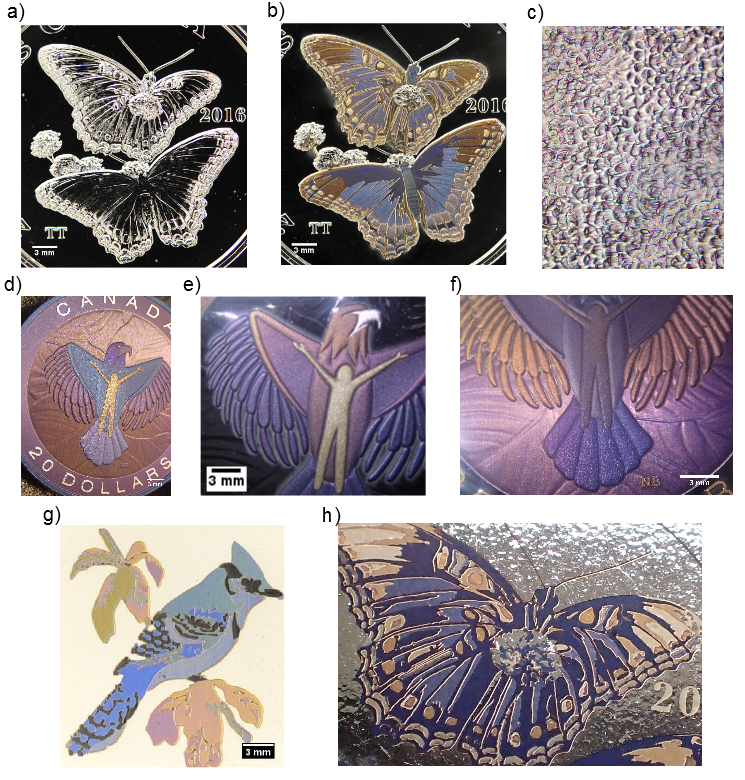}
\end{tabular}
\caption{Photographs of silver topographical butterfly coins (a) before and (b) after laser colouring; the topography has an overall height of $\sim$ 2 mm (surfaces as high as 4 mm were also coloured) without adjusting focus. The distinctive white areas on the butterfly are frosted ($\it{i.e}$., melted) finish; (c) optical microscope close up of the frosted regions, height of $\sim$ 1 to 2 $\mu$m. (d) Photograph of a silver eagle coin. The eagle is frosted and its surface is similar to that shown in (c). The difference in elevation is observed to create a colour gradient on the eagle wings (e) and tail (f). The eagles in (d-f) originate from different coins coloured using different laser parameters. (g) Blue Jay colouring on a flat silver coin, with permission from National Geographic. (h) Colouring of a butterfly, $\sim$ 10 mm wide, on a flat silver coin}
\label{fig2}
\end{figure} 
Previous work on picosecond and femtosecond laser colouring of metals \cite{Vorobyev2013a, Fan2014} only report a single set of laser parameters to obtain a given colour (less than 10 angle-independent colours). We show that there are in fact a very large number of laser parameter combinations that could be used to obtain a given Hue, as long as the total accumulated fluence remains the same -- as highlighted by the trend (master curve) in Fig. 1 (d) -- under the condition that the repetition rate remains fixed. Fig. 1 (e) shows a polar plot of Fig. 1 (d) with Hue plotted azimuthally and the logarithm of the total accumulated fluence plotted radially; a full 360$^{\circ}$ rotation in Hue is observed. The strength of the process and the master curve can also be seen in the aesthetic control of a single colour. For example, not only are we able to produce the colour blue, but we can also control the type of blue (\textit{e.g.}, navy blue, sky blue, etc...), as the Lightness of the colours, for the same Hue, scales up with laser fluence, $\phi$, as seen in Figs. 1 (b). The process allows us to generate thousands of colours, well-beyond what has been previously reported, enabling this technique to rival paint based processes currently used at the Royal Canadian Mint. Chroma, however, is observed to remain unaffected for each Hue when using different laser fluences, $\phi$. Figure 1 (f) is a CIE xy chromaticity diagram of the colours obtained via the different laser parameters in Fig. 1 (d). The rotation and overlapping of the data points on the diagram shows explicitly that the xy values of CIE XYZ colour space can be recovered for a fixed laser fluence, $\phi$, by simply adjusting the marking speed or line spacing. Deviations from the master curve in Fig. 1 (d), are attributed to the increasingly chaotic nature of the surface with increasing total accumulated fluence. While we are able to render reds reproducibly, there is less variety in the reds than for other colours due to the high slope in the red region on the master curve (Hue: 345 to 15). 	From equation (1), $v$, $L_s$ and $E$ can be changed independently to obtain a specific total accumulated fluence corresponding to a desired colour, Fig. 1 (d). \\

However, changing the repetition rate, $f$, with either $v$, $L_s$ and $E$ to obtain a desired total accumulated fluence resulted in colours other than what was expected based on the master curve of Fig. 1(d). To study the effect of the laser repetition rate on the colours, $f$ was changed proportionally with speed, $v$, to maintain a fixed total accumulated fluence with $E$ and $L_s$ kept constant. The colour palette obtained was similar to that shown in Fig. 1(a) but with fewer colours. Increasing $f$ and keeping it fixed while changing $v$, $L_s$ and $E$ would only create reduced colour palettes. The baseline in Fig. 1(d) increased with increasing repetition rate gradually cutting off the lower colours (\textit{i.e.}, blues, purples and reds). No colours other than yellow were observed at repetition rates above 400 kHz suggesting a local thermal accumulation effect. \\

Figs. 2 (a) and (b) show before and after images of the colouring process applied to silver coins of varying topography (up to 2 mm in height). The different colours were selected prior to laser colouring using the corresponding total accumulated fluence values from the master curve (Fig. 1 (d)). Precise colouring of the coin features was done using vision alignment software with pattern recognition. The vectors were obtained from an artist at the Royal Canadian Mint. The process is capable of uniform and viewing angle-independent colouring of elevated and complex surfaces that could not be coloured via traditional paint-based manufacturing processes. Moreover, compared to top-down fabrication techniques, this colouring process is efficient regardless of surface quality, allowing for the uniform colouring of frosted (\textit{i.e.}, melted) surfaces, Figs. 2 (a-f).\\

In addition, due to the dependence of the colours on the total accumulated fluence, colour gradients can also be produced by focusing the laser beam slightly above the surface, outside the confocal volume, making the colouring process sensitive to topography - see the eagle wings (e) and tail (f) in Fig. 2. The extent of the number of colours in the colour gradient is governed by the distance between the surface of the coin and the laser focus. At small distances (while still out of the confocal region) the colour gradient is observed to have a smooth transition in colours over a large area. Flat Figs. 2(g,h) and topographical (Fig. 2 (b)) surfaces situated within the confocal volume are, however, coloured uniformly. White and perfect black were also achieved using the appropriate laser parameter combinations, as shown in Fig. 2 (g).
\\
The long term stability of the colours was observed to depend on the colour. However, passivation coatings formed via 	atomic layer deposition on coloured silver surfaces protected the colours during aggressive humidity and tarnish tests carried 		out at the Royal Canadian Mint. The colours were, however, slightly red-shifted due to the passivation layer, an effect that 	can be pre-compensated by altering the write laser parameters. 
\subsection{Surface Analysis}
\begin{figure}[H]
\centering
\begin{tabular}{@{}p{1\linewidth}@{}}
\subfigimg[width=\linewidth]{\textbf{\textcolor{black}{}}}{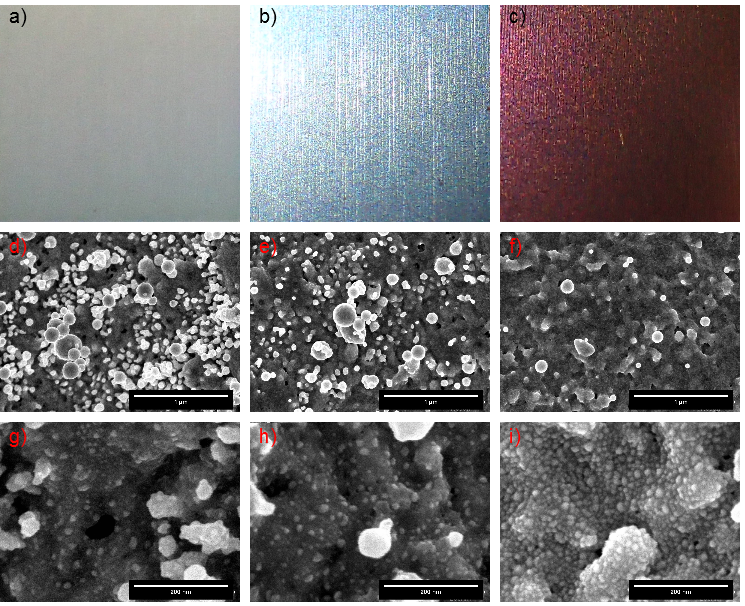} 
\end{tabular}
\caption{(a,b,c) Low-magnification optical microscope images of coloured surfaces; (d,e,f) low- and (g,h,i) high-magnification SEM images of corresponding surfaces. Surfaces processed using $\phi=1.12$ J/cm$^2$ at $v=11$ mm/s for (a,d,g) $L_s=5$ $\mu$m (Hue = 216.5, Cyan), (b,e,h) $L_s=10$ $\mu$m (Hue = 269.3, Blue), and (c,f,i) $L_s=30$ $\mu$m (Hue = 17.2, Red). The number of medium NPs is observed to decrease with increasing $L_s$ (d,e,f -- scale 1 $\mu$m), and the number of small NPs to increase with $L_s$ (g,h,i -- scale 0.2 $\mu$m).}
 \label{fig3}
\end{figure}

Extensive SEM analyses of regions exhibiting different colours on silver reveal 3 distinct classes of particles differentiated by size: large ($r\geq75$  nm), medium ($10.7\leq r$ $\textless$ 75 nm) and small ($r$ $\textless$ 10.7 nm) where $r$ is the radius of a particle. 
These particle classes were obtained from statistical analysis of SEM images which produced histograms with well-defined ranges of particle sizes (see Fig. S. 4(a,b,c), supplementary material).

Figs. \ref{fig3}(a-c) show three coloured surfaces with corresponding low- and high-magnification SEM images. The particle density is found to change significantly with line spacing, as shown in Figs. \ref{fig3}(d,g). 
%for $L_s=5$ $\mu$m, (e,h) for $L_s=10$ $\mu$m and (f,i) for $L_s=30$ $\mu$m for silver exposed to a fluence $\phi=1.12$ J/cm$^2$ at $v=11$ mm/s. 
In Figs. \ref{fig3}(d-i), the small and medium particles are seen to form random networks with the medium-sized particles sparsely covering the surface and the small ones more densely distributed across the irradiated region. The formation of NPs is believed to come from the combination of thermal effects\cite{Che2008, Galhenage2013, Roque2005, Fazio2014} and the re-deposition of particles following laser ablation\cite{Perriere2007, Tillack2004a, Tillack2004}.

%Figures S. 1(a,b,c) (supplementary materials) show histograms of the number of particles on the surface over an area of 1 $\mu m^{2}$ versus particle radius for $L_s=5$, 10 and 30 $\mu$m, respectively. Two discernible bumps (bimodal distribution) are noted in the histograms, corresponding to the small and medium particles. 
Upon close examination of the SEM images, it appears that the small particles are in reality approximately hemispherical, an observation supported by a cross-section of a coloured silver sample obtained through focused ion beam milling and imaging. The small particles can be viewed as spheres partially embedded into the silver surface.
\\
SEM analyses of coloured regions reveal that the number density of the small particles, produced using different laser parameters, follows its own clear distinctive trend with total accumulated fluence, (Fig. S. 3 (a), supplementary materials), similar to that of Fig. 1 (d), whereas the number density of the medium particles does not (Fig. S. 3 (b), supplementary materials). This observation suggests that the small particles play a major role in the colours perceived, even though they have not been considered in previous works. The mean radius of small and medium particles was found to remain approximately constant as a function of line spacing (Figs. S. 3 (c,d), determined from the analysis of three SEM images per line spacing or colour). However, the mean inter-particle distance (wall-to-wall) changes with line spacing, Figs. S. 3 (e,f), suggesting that the colours are affected by near-field interactions between nanoparticles in close proximity \cite{Rechberger2003,Romero2006,Jain2010,Liz2006},  particularly the associated surface plasmon resonance frequency \cite{Liz2006}.
\\
Wavelength-dispersive spectroscopy (WDS) analysis of the different colours in Figure 1, showed no difference in the amount of oxidation measuring 2.8 $\pm$ 0.4$\%$ oxygen content for all Hue values tested. Monte Carlo simulations (WinXray) of silver oxide layers, under the same WDS operating conditions, gave an equivalent oxide thickness of $\sim$ 2 nm. The same conclusion was gathered from the invariability of the oxygen content in separate energy dispersive spectroscopy (EDS) and X-ray photoelectron spectroscopy (XPS) measurements of the different coloured surfaces. The XPS analyses of the coloured surfaces also showed no sulphur content.

\subsection{FDTD Simulations}
Our detailed numerical simulations show that colour formation can be viewed as a selective absorption process occurring due to plasmonic resonances. White light incident on a nanostructured metallic surface is not fully reflected, as it would be for a smooth surface. Rather, narrow-band spectral components (colours) are subtracted due resonant and absorptive processes in nanoclusters (homogeneous and/or heterogeneous), single particles, lattice resonant modes, and absorption in the medium, consistent with another study \cite{Ng2015}. What survives is reflected (there is no transmittance).
\begin{figure}[H]
\centering
\begin{tabular}{@{}p{.8\linewidth}@{}}
\subfigimg[width=\linewidth]{\textbf{\textcolor{black}{a)}}}{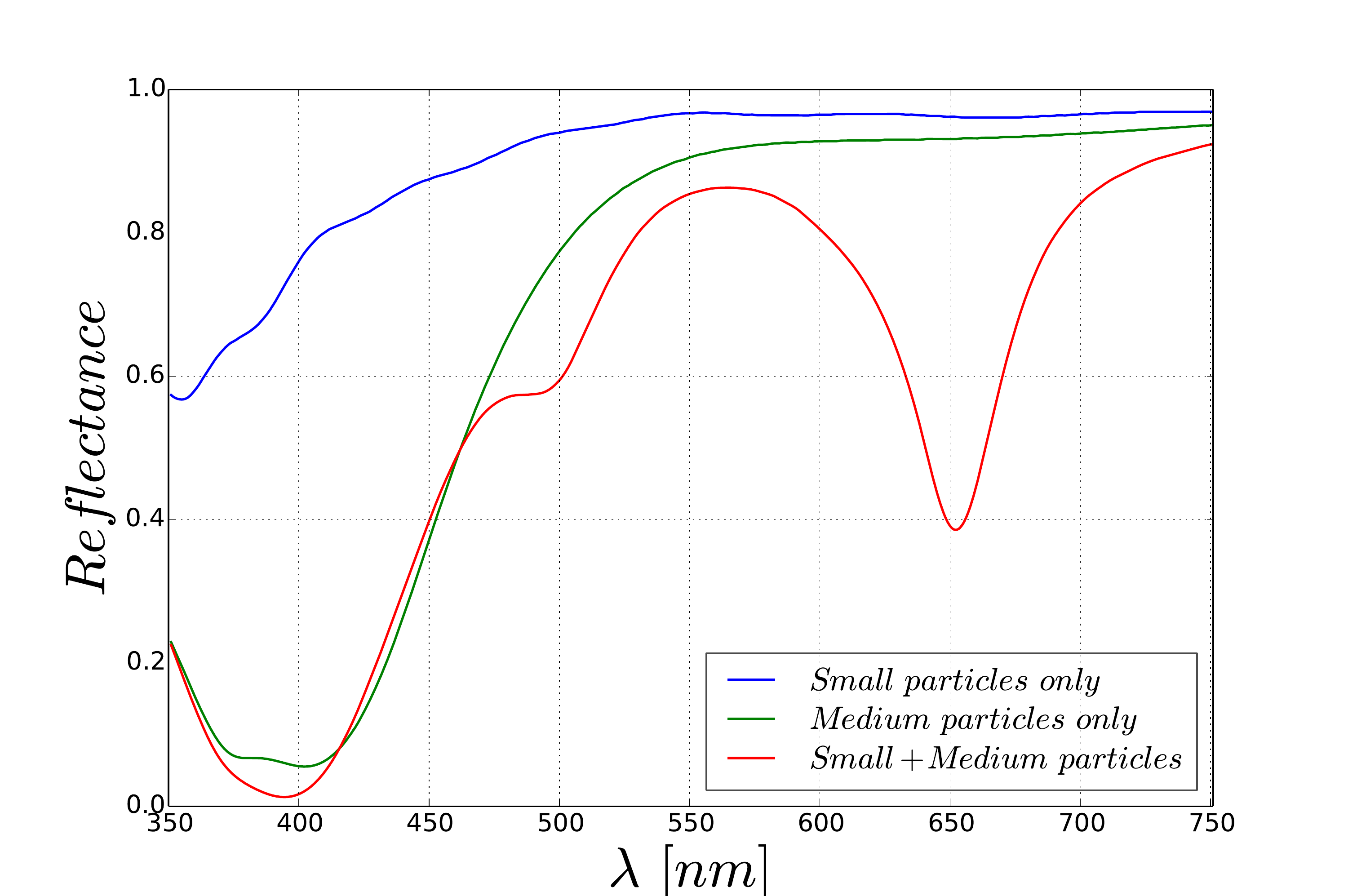}\\
\subfigimg[width=\linewidth]{\textbf{\textcolor{black}{b)}}}{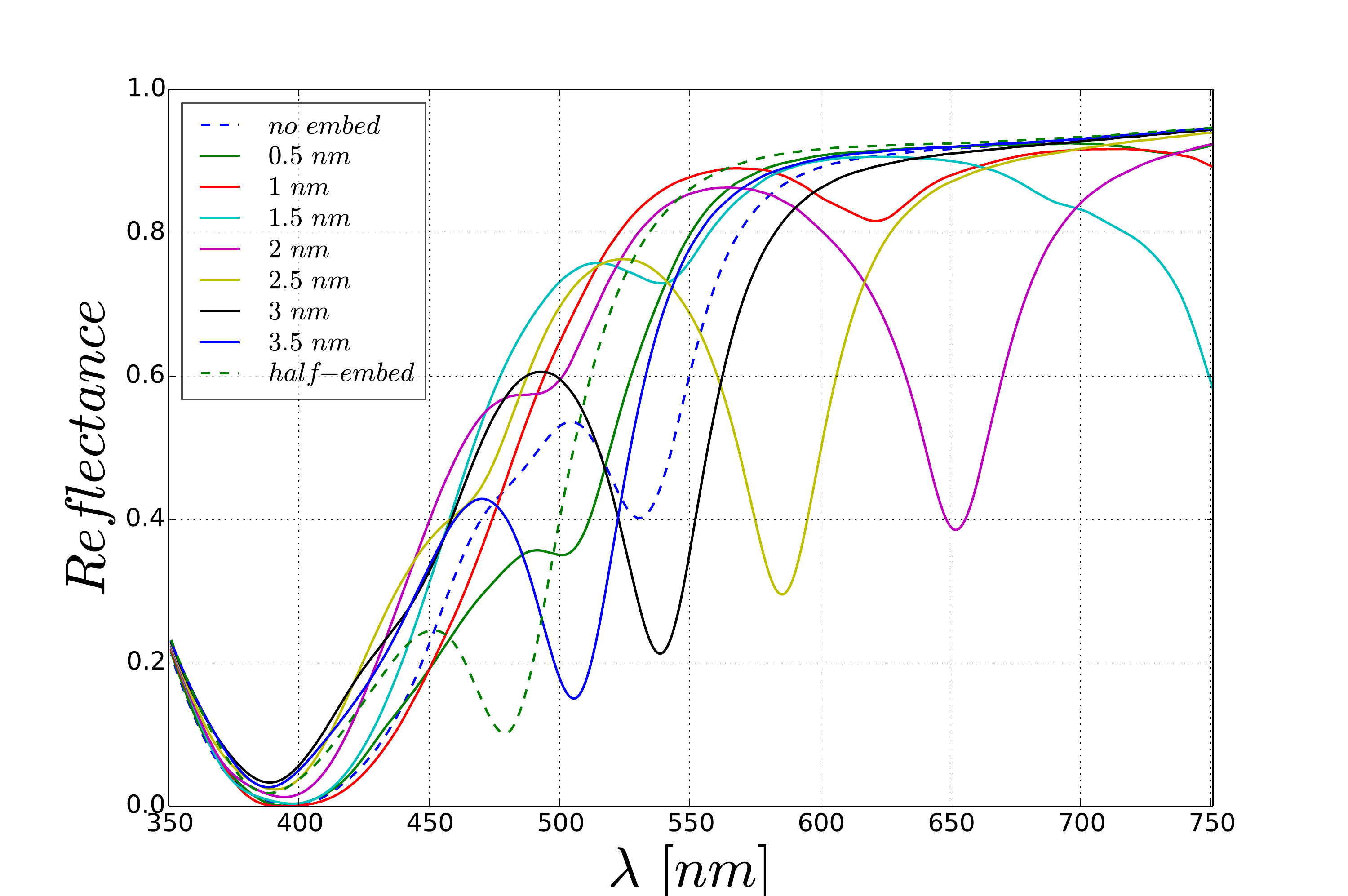}
\end{tabular}
\begin{tabular}{@{}p{.7\linewidth}@{}}
\subfigimg[width=\linewidth]{\textbf{\textcolor{black}{c)}}}{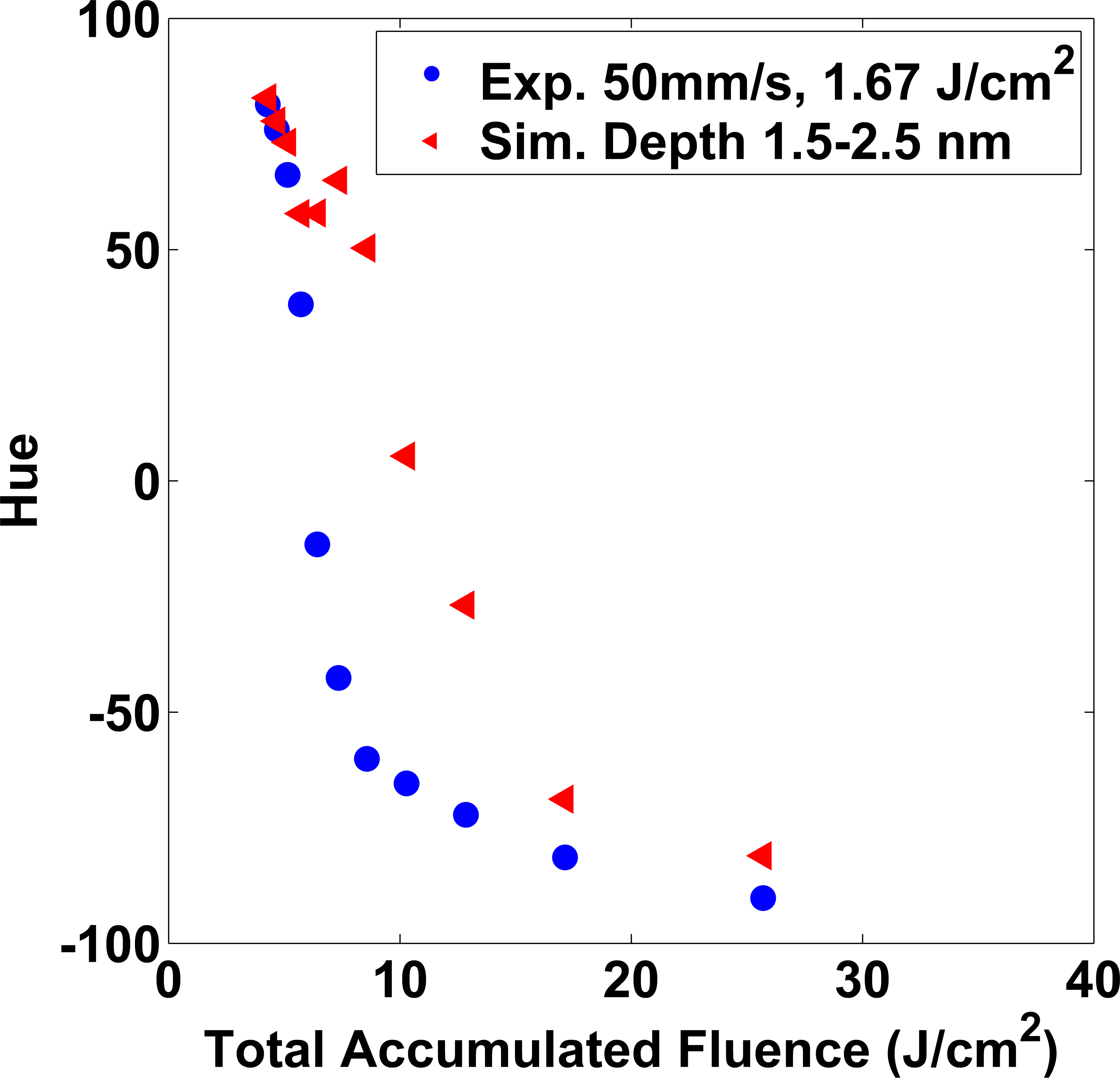}
\end{tabular}
\caption{(a) Computed reflectance spectra considering only small NPs embedded by $R_s$/2, only medium NPs embedded by $R_m$/2, and medium and small NPs embedded by $R_m$/2 and $R_s$/2, respectively. (b) Reflectance spectra computed by varying the embedding of the small NPs in steps of 0.5 nm (the medium NPs are embedded $R_m$/2). (c) Graph of Hue values vs. total accumulated fluence comparing the measured and computed values for small NPs embedding of 1.5-2.5 nm, and medium NPs not embedded.}
\label{fig5}
\end{figure} 
The analysis of SEM images produced statistics of the distributions, \textit{i.e.}, average radii and average inter-particle distances for small, medium and large NPs. We carried out simulations considering only small and medium Ag NPs uniformly distributed on a Ag substrate. This is supported by the two discernible bumps (bimodal distribution) noted in the histograms (Fig. S. 4 (a,b,c), supporting information). The large particles are neglected due to their low density. The radii of small and medium sized NPs are $R_s$ and $R_m$, and the inter-particle distances (centre-to-centre) are $D_s$ and $D_m$, respectively.
In each case the NPs were periodically arranged in a hexagonal configuration, which produces a hexamer unit cell. The hexamer has $D_{6h}$ symmetry, and based on group theory, exhibits polarization independence (isotropy), as verified by simulations \cite{Hent2010}. We chose $D_m$ as an integer multiple of $D_s$ to allow the application of periodic boundary conditions (PBCs). The periodicity of the unit cell is small enough to preclude coupling by diffraction into surface plasmon waves on the Ag surface.
\\
By changing the surface density of small and medium sized NPs, their radii, and their level of embedding into the substrate, several types of heterogeneous nanoclusters can be formed. Embedding transforms the NPs gradually from spherical to hemispherical. A nanocluster has a central nanoparticle (cNP) with nearby nanoparticles arranged in a ring (rNPs). Plasmonic nanoclusters embedded in a homogeneous medium or on a dielectric have been extensively studied (dimer, trimer, quadrumer, tetramer, hexamer, heptamer)\cite{Luk2010,Mirin2009,Hent2010}, but not on a metal substrate, and not arranged and embedded as inspired by SEM images of laser-processed metal surfaces. Plasmonic nanoclusters exhibit resonances which can lead to selective absorption\cite{Mirin2009}. 
We have observed different types of resonant modes in our simulations; we will refer to them as cluster (or collective) resonances.

In the simulations that follow, we consider the case of line spacing $L_s=5$ $\mu$m marked at $v=50$ mm/s with a laser fluence $\phi=1.67$ J/cm$^2$, having statistical average parameters $R_s=4$ nm, $R_m=34.3$ nm, $D_s=13.5$ nm and $D_m=108$ nm.

%Fig5
In Fig. 4 (a) we show the computed reflectance spectrum considering small NPs only, medium NPs only, and the combination of both NP sizes; the NPs are half-embedded, \textit{i.e.}, embedded by half their radius. The medium size NPs are observed to produce colours in the absence of the small ones, but changing $D_m$ and the level of embedding does not reproduce the range of colours observed experimentally. Alternatively, considering only small NPs produced colours that are too light to be discernible. However, the interaction between small and medium NPs drastically alters the computed reflection spectrum, as noted by the appearance of a deep dip at $\lambda$ = 650 nm in Fig. 4 (a), which is due to a new cluster resonance. This will produce a discernible final colour. In our simulations we observed a very high sensitivity to embedding, which changes dramatically the geometry of the nanoclusters, and consequently, their resonance characteristics. In particular, the embedding of the small NPs increases the size of the nano-gaps between the cNP and the rNPs. In Fig. 4 (b) we show the reflectance by embedding the small NPs in steps of 0.5 nm. A blue-shift in the cluster resonance is observed when the embedding of small NPs is increased. A similar blue-shift due to cluster expansion was reported in \cite{Hent2010}.
In Fig. 4 (b) we also see other smaller dips in the reflectance curves. These are due to nanocluster resonances arising from aggregates of small NPs. In Fig. 4 (c) we compare the experimental results with the FDTD simulations. A qualitative agreement was found for medium NPs not embedded and small NPs embedded into the substrate by 2.5 nm. We were able to reproduce by simulations the full colour palette, supporting the role of plasmons in the colour rendition. We found that a large set of geometries can produce the same reflectance, in the same way as a large set of laser parameters can produce the same colour.

%DFT field distribution

\begin{figure}[H]
  \centering
  \begin{tabular}{@{}p{.8\linewidth}@{}}
	\subfigimg[width=\linewidth]{\textbf{\textcolor{white}{}}}{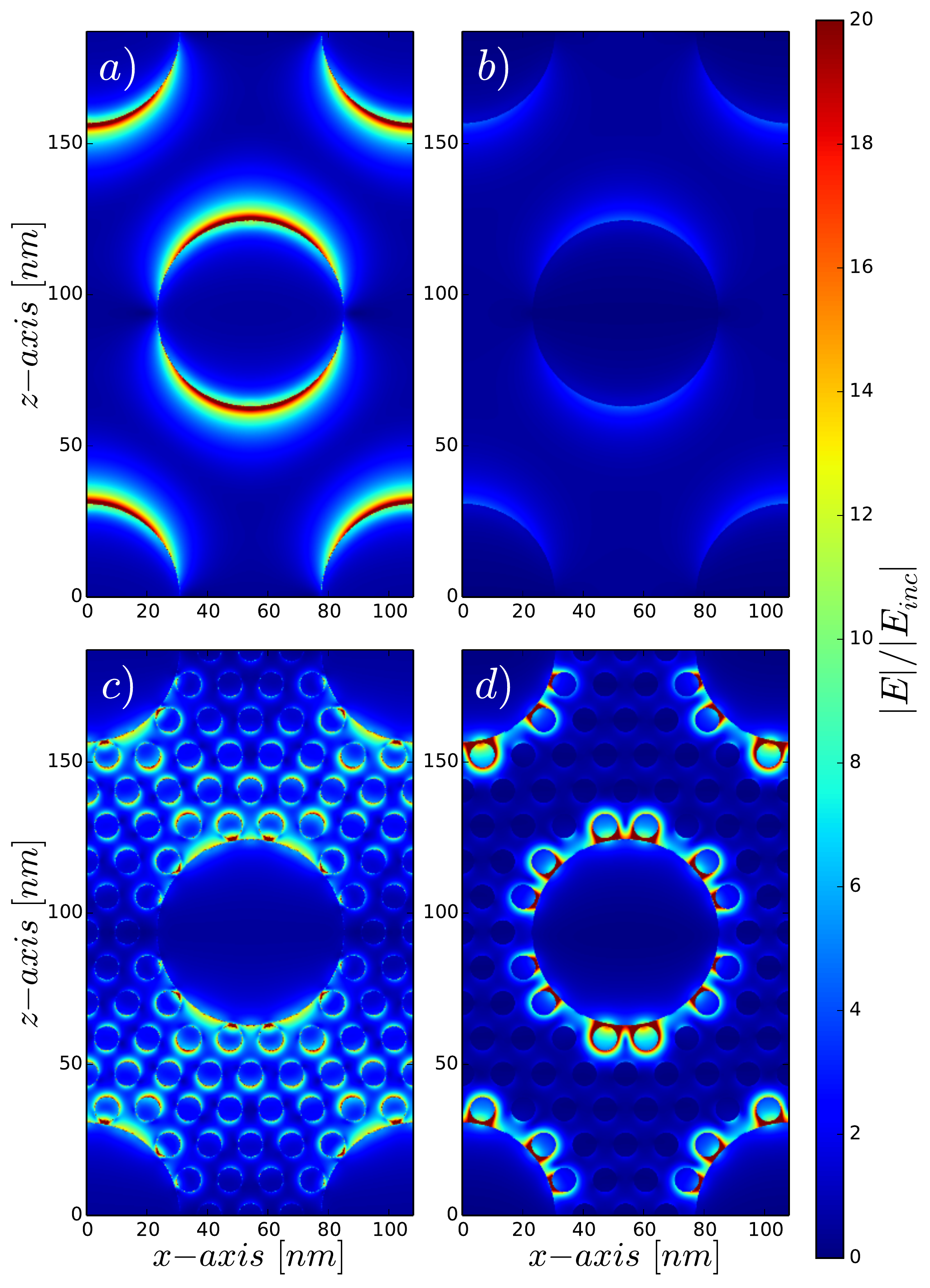}
	\end{tabular}
	\begin{tabular}{@{}p{.8\linewidth}@{}}
	\subfigimg[width=\linewidth]{\textbf{\textcolor{black}{e)}}}{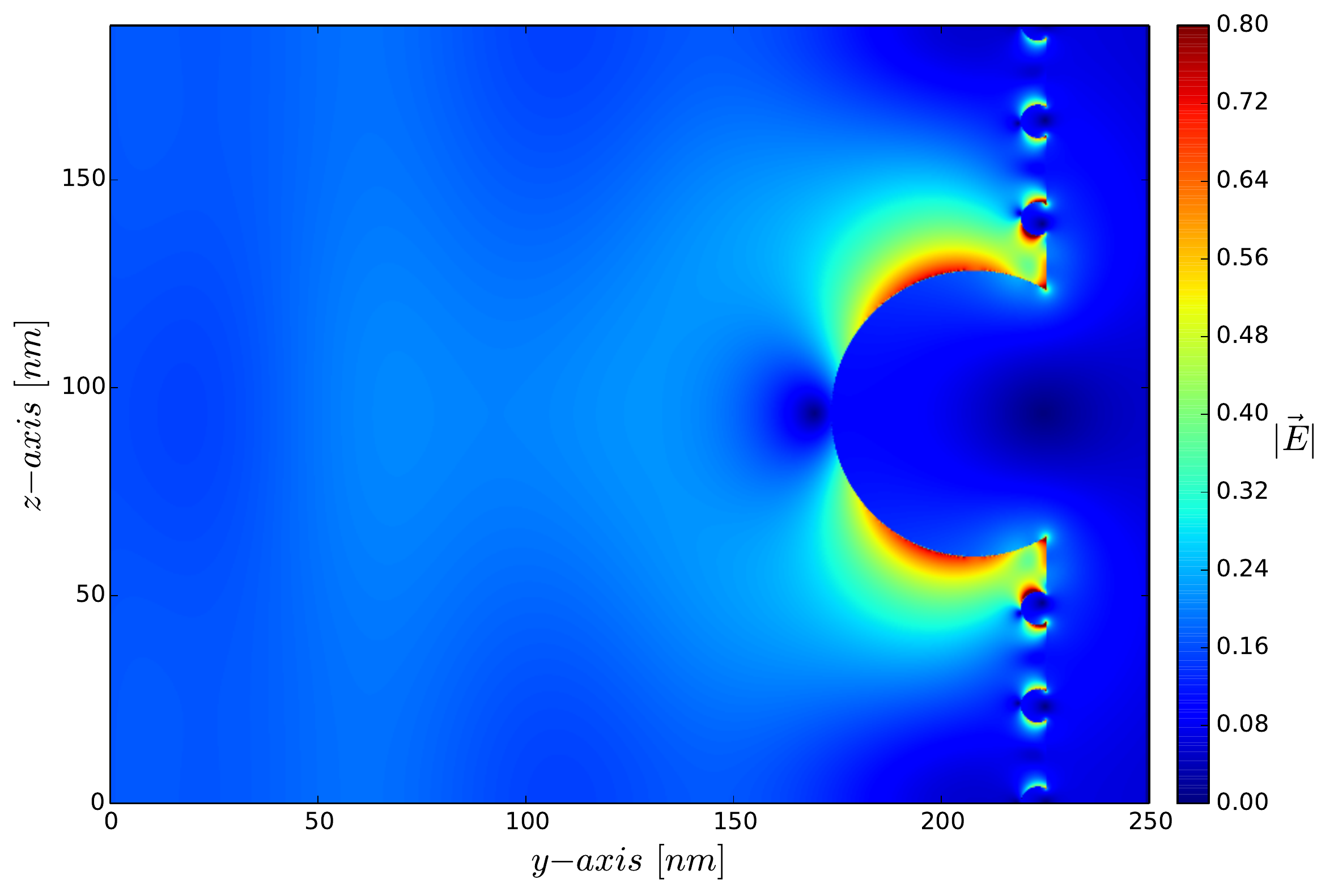}
  \end{tabular}
	\begin{tabular}{@{}p{.7\linewidth}@{}}
	\subfigimg[width=\linewidth]{\textbf{\textcolor{white}{f)}}}{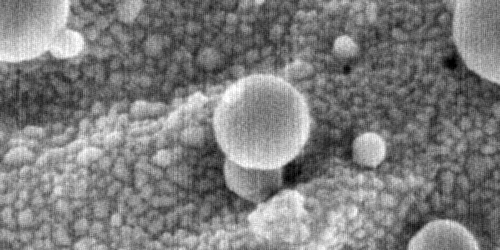}
  \end{tabular}
  \caption{FDTD simulations showing the electric field distribution for only medium size NPs at (a) $\lambda$ = 390 nm and (b) $\lambda$ = 650 nm, and for medium and small NPs at (c) $\lambda$ = 390 nm and (d) $\lambda$ = 650 nm. (e) Time-domain snapshot of medium and small nanoparticles embedded by $R_{m}$/2 and $R_{s}$/2, respectively. (f) SEM image for $\phi$=1.67 J/cm$^2$ at $v=50$ mm/s with $L_s=8$ $\mu$m.}
  \label{fig6}
\end{figure}

In Figs. \ref{fig6}(a-d) we show the electric field distribution at the free-space optical wavelengths of $390$ nm (a,c) and $650$ nm (b,d), which are the wavelengths at which we observe absorption dips in Fig. \ref{fig5}(a).
We show $xz$ planes cut 2 nm above the silver surface for medium NPs only (a,b), and medium and small NPs (c,d). 
In Fig. \ref{fig6}(a) we observe the resonance of the medium sized NP at $390$ nm, as predicted by Mie theory for the same sphere in air.
In our case, the resonance produces a dip in the reflectance due to the presence of the substrate.
Fig. \ref{fig6}(b) shows that the medium NPs only do not produce any absorption at $650$ nm, and the reflectance is very high. 

In Fig. \ref{fig6}(c) we observe that the near-field intensity of the medium NPs at $390$ nm is reduced by the presence of the small NPs, which act as plasmonic chain waveguides \cite{Maier2002}, coupling the medium size NPs.
\\
In Figs. 5 (a-d) we show the electric field distribution at the free-space optical wavelengths of $\lambda$ = 390 nm (a,c) and $\lambda$ = 650 nm, which are the wavelengths at which we observe absorption dips in Fig. 4 (a). We show xz planes cut 2 nm above the silver surface for medium NPs only (Figs. 5 (a,b)), and medium and small NPs (Figs. 5 (c,d)).  In Fig. 5 (a) we observe the resonance of the medium sized NPs at $\lambda$ = 390 nm, as predicted by Mie theory for the same sphere in air. In our case, the resonance produces a dip in the reflectance due to the presence of the substrate. Fig. 5 (b) shows that medium NPs only do not produce any absorption at $\lambda$ = 650 nm, and the reflectance is very high. In Fig. 5 (c) we observe that the near-field intensity of the medium NPs at $\lambda$ = 390 nm is reduced by the presence of the small NPs, which act as plasmonic waveguides \cite{Maier2002}, coupling the medium size NPs. When small and medium NPs are illuminated at $\lambda$ = 650 nm (Fig. 5(d)), a heterogeneous cluster resonance is excited, producing field enhancement in the nano-gaps, and ultimately the strong absorption observed in Fig. 5 (a). In Fig. 5 (e) we show a snapshot of the time-domain simulation in a yz plane cut through the centre of the medium NPs. The plane wave excitation pulse has just hit the nanostructures inducing localized surface plasmons in all NPs which are clearly visible. The time-evolution of the excitation in the xz and yz planes is shown in Movies S. 1 and S. 2 (supporting information), respectively.
In Fig. 5 (f) we show an SEM image for the case of $\phi$ = 1.67 J/cm2 at $v$ = 50 mm/s with $L_s$ = 8 $\mu$m, which further justifies our simulation approach. We observe the presence of two sets of particles (small and medium sizes) having a random distribution. In our simulations we considered perfect periodicity of the NPs, which gives isotropy with respect to the incident polarization, narrowband resonances, and ultimately colour selectivity. This can be considered a good qualitative approximation. In \cite{Nishijima2012} random clusters were investigated, and the authors reported a broadening of the resonances. This matches our experiments which show broader spectra with respect to simulations.

\section{Conclusion}
We described a universal and deterministic process for colouring noble metals with and without topography. Each individual colour can be linked to a total accumulated fluence. The colours originate from random distributions of small and medium nanoparticles embedded into the surface, induced and controlled by laser exposure.  The randomness of the nanoparticle networks is modelled effectively by assuming a periodic structure defined by statistical averages of nanoparticle size and separation. We have demonstrated that plasmonic effects arising in heterogeneous nanoclusters explain the full palette of experimental colours. The small nanoparticles in particular, which have always been neglected in post-surface analyses following laser exposure, play a fundamental role in the colour formation. In fact, they change the geometry of the nanoclusters, the cluster resonance, and ultimately the perceived colour. The new proposed method decreases the colouring time of large metal surfaces, and opens the door to large-scale industrial applications for anti-counterfeiting, bio-sensing, bio-compatibility, and the decoration of consumer products such as jewels, art, architectural elements and fashion items.

\section{Methods}
In our experiments, 1064 nm light from a 15 W Duetto (Nd:YVO$_{4}$, Time-Bandwidth Product) mode-locked MOPA laser, operating at a repetition rate of 50 kHz and producing 10 ps pulses, was focused on the metal surface using an F-theta lens (f=163 mm, Rodenstock). The pulse energy for the presented results ranged between 3.4 and 91.4 $\mu$J. The laser was fully electronically integrated and enclosed by a third party for industrial applications (GPC-PSL, FOBA). For accurate focusing, the surface of the samples was located using a touch probe system. The silver samples were of 99.99$\%$ purity and not polished prior to machining to meet requirements of reproducibility in industrial applications. For machining, the samples were placed on a 3-axis translation stage with a resolution of 1 $\mu$m in both the lateral and axial directions. The samples were raster scanned using galvanometric XY mirrors (Turboscan 10, Raylase) displacing the beam in a top to bottom fashion with a mechanical shutter blocking the beam between successive lines. The laser polarization was parallel to the marking direction. The laser power was computer controlled via a laser interface and calibrated using a power meter (3A-P-QUAD, OPHIR). A spot size of 14 $\mu$m was obtained from semi-logarithmic plot of the square diameter of the modified region, measured with a scanning electron microscope (SEM), as a function of energy, following the procedure described in \cite{Jandeleit1996}. High resolution SEM (JSM-7500F FESEM, JEOL) images were obtained using secondary electron imaging (SEI) mode. Colours were quantified using a Chroma meter (CR-241, Konica Minolta) with the CIELCH colour space, 2 observer and illuminant C (North sky daylight); where L is colour Lightness, C is Chroma (colour saturation) and H is Hue (colour value associated with a 360 polar scale). For analysis of the SEM images, a Matlab program was  written to locate the position of each particle and record its diameter and the wall-to-wall inter-distance spacing to its nearest-neighbours. 

Three dimensional finite-difference time-domain (FDTD) simulations\cite{Taflove2005, Taflove2013} have been performed to determine the origin of the colour formation process. We used in-house 3D-FDTD parallel code \cite{Lesina2015,Vaccari2014} on an IBM BlueGene/Q supercomputer (64k cores) part of the Southern Ontario Smart Computing Innovation Platform (SOSCIP).
 
The nanoparticles are arranged on the $xz$-plane. The system is excited by a $z$-polarized plane wave. This is a broadband electromagnetic pulse propagating along the $y$-direction from air and impinging on the nanostructured surface. 
The analysis was performed over the wavelength range 350-750 nm in a single run of the code by in-line discrete Fourier transform (DFT). A space-step of $0.25$ nm was used for the simulations, and $0.125$ nm for visualization quality (Figs. \ref{fig6}(a-e) and Movies). 
The dispersion of silver was introduced by the Drude+2CP model \cite{Vial2011}. This model was implemented in FDTD by the auxiliary differential equation (ADE) technique \cite{Prokopidis2013}. The simulation domain in the direction of the plane wave propagation is truncated by convolutional perfectly matched layers (CPML) absorbing boundary conditions \cite{Roden2000}. The simulations required up to 16k cores. 
The theoretical reflectance spectrum is calculated by integration of the Poynting vector in the backward far-field region. 
The experimental reflectance spectra (colours) were reconstructed using an in-house Matlab code, weighting each frequency composing the spectra to the spectral sensitivity of the eye.

\appendix
\subsection*{Acknowledgements}
We acknowledge the Royal Canadian Mint, the Natural Sciences and Engineering Council of Canada, the Canada Research Chairs program, the Southern Ontario Smart Computing Innovation Platform (SOSCIP), and SciNet. We would like to acknowledge Alessandro Vaccari at Fondazione Bruno Kessler (Italy), Josh Baxter and Meagan Ginn COOP students at the University of Ottawa. 
\subsection*{Authors contributions}
%F.L., C.E.-L., S.N.S., K.V., and H.B. designed and synthesized the compounds. F.S., and E.L. set up and performed the experiments and the data analysis. G.K. and R.S. carried out the calculations. F.S., G.K., K.V., H.B., R.S. and E.L. wrote the paper. All authors discussed the results and commented on the manuscript.
J-M.G. and G.C. developed the laser colouring technology. 
A.W. proposed and directed the project, and supervised the experiments.
L.R. and P.B. co-directed the study and supervised the simulations.
J-M.G., G.C. and M.C. conducted the laser colouring experiments and the statistical analysis of the surfaces.
G.C. wrote the Matlab codes.
A.C.L., J-M.G., L.R. and P.B. discussed the simulation strategy and the simulation results. 
A.C.L. setup the FDTD code, conducted the simulations and generated the movies.
A.C.L., L.R. and P.B. formulated the theoretical conclusions.
J-M.G. and A.C.L. co-wrote the paper and prepared the figures.
L.R., P.B. and A.W. corrected and commented on the manuscript.
All authors discussed about experimental results.

%J-M.G. developed the technology, conducted the laser colouring experiments, SEM images acquisition and statistical analysis of surfaces, and wrote the manuscript.
%A.C.L. developed the simulation strategy, wrote the FDTD code, conducted all the simulations, and wrote the manuscript.
%G.C. conducted the laser colouring experiments, SEM images acquisition and statistical analysis of surfaces, wrote the Matlab code.
%M.C. conducted the laser colouring experiments, SEM images acquisition and statistical analysis of surfaces.
%L.R. codiscussed the results, provided theoretical support and corrected the manuscript.
%P.B. discussed the results, provided theoretical support and corrected the manuscript.
%A.W. supervised the experiments and corrected the manuscript.
\subsection*{Competing financial interests}
The authors declare that they have no competing financial interests.
%\subsection*{Correspondence}
%Correspondence and requests for materials
%should be addressed to Jean-Michel Guay~(email: guay.jeanmichel@gmail.com).

\bibliography{references}{}
\end{multicols}
\end{document}